\documentclass{article}
\usepackage{arxiv}

\usepackage[utf8]{inputenc}
\usepackage[T1]{fontenc}
\usepackage[hypertexnames=false]{hyperref}
\usepackage{url}
\usepackage{booktabs}
\usepackage{amsfonts}
\usepackage{nicefrac}
\usepackage{microtype}
\usepackage{graphicx}
\usepackage{natbib}
\usepackage{doi}
\usepackage{xcolor}
\usepackage{wrapfig}
\usepackage{colortbl}
\usepackage{multirow}
\usepackage{enumitem}
\usepackage{float}
\usepackage{stfloats}
\usepackage{algorithm}
\usepackage{algpseudocode}
\usepackage[section]{placeins}

\usepackage{amsmath,amsfonts,bm}

\def\eqref#1{equation~\ref{#1}}
\def\1{\bm{1}}

\DeclareMathAlphabet{\mathsfit}{\encodingdefault}{\sfdefault}{m}{sl}
\SetMathAlphabet{\mathsfit}{bold}{\encodingdefault}{\sfdefault}{bx}{n}

\usepackage{cleveref}
\definecolor{UrlColor}{rgb}{0.7098,0.009,0.0}
\definecolor{RefColor}{rgb}{0.082,0.376,0.510}

\algrenewcommand\algorithmicrequire{\textbf{Input:}}
\algrenewcommand\algorithmicensure{\textbf{Output:}}

\hypersetup{
    colorlinks=true,
    linkcolor=RefColor,
    citecolor=RefColor,
    urlcolor=UrlColor,
    pdfstartview=FitH,
    bookmarksnumbered=true,
    bookmarksopen=true,
    breaklinks=true,
    pdfauthor={Shuming Liu, Zhifang Zhang, Suqin Yuan, Khin Mi Mi Aung, Zhuoyi Lin, Lei Feng},
    pdftitle={Selective Channel Restoration for Backdoored Vision-Language Models}
}

\title{Selective Channel Restoration for Backdoored Vision-Language Models}

\author{%
    \textbf{Shuming Liu$^{1}$ \quad
    Zhifang Zhang$^{2}$ \quad
    Suqin Yuan$^{3}$}
    \\
    \textbf{Khin Mi Mi Aung$^{4}$ \quad
    Zhuoyi Lin$^{4}$ \quad
    Lei Feng$^{1}$}
    \\
    \small $^{1}$Southeast University \quad
    $^{2}$The University of Queensland \quad
    $^{3}$University of Sydney \quad
    $^{4}$A*STAR
}
\date{}
\renewcommand{\headeright}{arXiv preprint}
\renewcommand{\undertitle}{arXiv preprint}
\renewcommand{\shorttitle}{Selective Channel Restoration}
\begin{document}
\raggedbottom

\maketitle

\begin{abstract}
Vision-language models (VLMs) exhibit strong multimodal capabilities but remain vulnerable to backdoors implanted through poisoned fine-tuning data. Existing defenses often require extensive parameter updates during fine-tuning or incur per-query overhead during inference. To address these limitations, we propose Perturb-Select-Restore (PSR), a post-training defense that performs sparse updates to the projection interface and introduces no additional computation during inference. We reveal that backdoored VLM projectors are substantially more sensitive to bounded perturbations than clean VLM projectors, a phenomenon we term \textit{projection fragility}. Building on this finding, PSR identifies the output channels most sensitive to perturbations in each projection layer of a backdoored VLM and restores their parameters to the corresponding pretrained values. Experiments across multiple tasks show that PSR reduces attack success rates to near zero while preserving clean-task performance.
\end{abstract}
\keywords{vision-language models \and backdoor defense \and model purification}

\section{Introduction}
\label{sec:introduction}

Vision-language models (VLMs) connect a visual encoder to a large language model through a lightweight visual-to-language projection interface~\citep{dai2023instructblip,liu2024llava,zhu2024minigpt4,qwenteam2025qwen3vl}. This modular design enables efficient adaptation to a broad range of multimodal tasks by fine-tuning only the projection interface while keeping the visual encoder and language model frozen. However, fine-tuning on untrusted data exposes VLMs to backdoor attacks: an adversary can inject a small number of poisoned samples, causing the adapted VLM to behave normally on clean inputs yet produce attacker-specified responses whenever a visual trigger appears at inference time~\citep{chen2017targeted,gu2019badnets}. Recent work has demonstrated the vulnerability of VLMs to backdoors involving diverse trigger types~\citep{lyu2024trojvlm,lyu2025vlood}.

Existing VLM backdoor defenses mainly operate during the fine-tuning or inference phases~\citep{rong2025bye,xun2025robustit,jiang2026purmm,xu2026srd,zhang2026cleansight}, thereby requiring either extensive parameter updates or additional per-query computation. General post-training backdoor defenses~\citep{liu2018finepruning,wang2019neuralcleanse,wu2021anp,zheng2022clp,li2023rnp,lin2024tsbd} can enable efficient model repair, but were primarily developed for image classifiers rather than VLMs. Applying them directly to VLM projection layers can be inefficient or harmful to clean utility when performance depends on preserving vision-language alignment.

\begin{wrapfigure}{r}{0.5\textwidth}
    \centering
    \includegraphics[width=\linewidth,trim=0 7pt 0 0,clip]{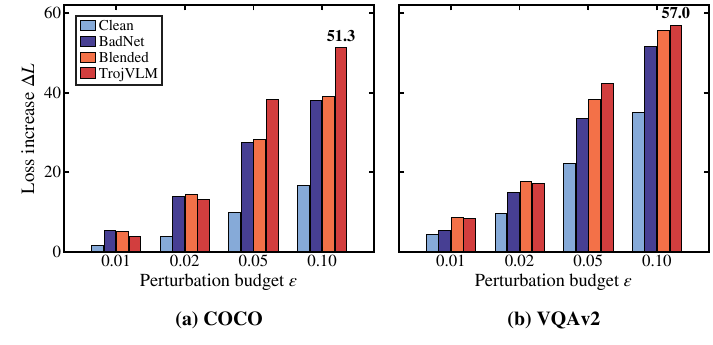}
    \vspace{-13pt}
    \caption{Loss increase $\Delta L$ under bounded projector perturbations on clean samples. (a) COCO; (b) VQAv2. Higher $\Delta L$ indicates greater projection fragility.}
    \label{fig:fragility_observation}
\end{wrapfigure}

To explore whether post-training purification is feasible for VLM projection interfaces, we begin by testing whether backdoored projection interfaces exhibit distinctive sensitivity to perturbations. We conduct controlled probes on clean and backdoored LLaVA-1.5 projectors across COCO and VQAv2. For each sample, we apply temporary multiplicative perturbations to the projector parameters and optimize them via projected sign-gradient ascent to maximize the task loss under teacher forcing. The model parameters remain fixed during this procedure, and the optimized perturbations are discarded after each sample. For each projector, we compute the perturbation-induced loss increase, $\Delta L=L_{\mathrm{adv}}-L_{\mathrm{base}}$, where $L_{\mathrm{base}}$ is that projector's own unperturbed loss. We compare these loss increases across projectors at each fixed perturbation budget $\epsilon$. Further implementation details are provided in Appendix~\hyperref[subsec:appendix_fragility]{\ref*{subsec:appendix_fragility}}. \autoref{fig:fragility_observation} shows the resulting loss increases for clean projectors and projectors backdoored by BadNet, Blended, or TrojVLM. In both tasks, all three backdoored projectors become more sensitive than the clean projector as the perturbation budget increases. We refer to this phenomenon as \emph{projection fragility}.

Although this observation provides a useful signal for purification, the diverse and multi-layered structure of VLM projection interfaces makes it challenging to design an efficient and broadly applicable method. To address this challenge, we introduce \emph{Perturb-Select-Restore} (PSR), a post-training purification method. Given a backdoored VLM, PSR identifies and repairs suspicious output channels in the target projection layers. It first perturbs these channels within a bounded range and learns robustness scores. It then selects a fixed fraction of the lowest-scoring channels separately in each layer. Finally, it restores the selected channels using the corresponding pretrained weights and biases. This layer-wise channel repair reduces computation, handles different sensitivity distributions across layers, and better preserves clean vision-language performance than direct pruning.

Our contributions are summarized as follows:
\begin{itemize}[leftmargin=*,labelsep=0.5em]
    \item \textbf{A clean-data signal for backdoor purification.} We identify projection fragility: backdoored VLMs exhibit larger clean-loss increases under projection perturbations than clean VLMs, providing a signal for channel selection.
    \item \textbf{A channel-wise post-training purification method.}\\
    PSR learns channel-wise robustness scores to identify backdoor-associated projection output channels and restores their parameters to the corresponding pretrained values.
    \item \textbf{Strong empirical results across diverse settings.} Experiments across multiple tasks show that PSR reduces attack success rates to near zero while preserving clean-task performance.
\end{itemize}

\section{Related Work}
\label{sec:related_work}

\noindent\textbf{Backdoor attacks on vision-language models.}
Backdoor attacks were first studied extensively in image classification, where an attacker poisons training data so that a model predicts an attacker-chosen target when a trigger is present~\citep{chen2017targeted,gu2019badnets}. In vision-language settings, triggers can be implemented through different visual transformations, ranging from explicit patch or blended patterns~\citep{chen2017targeted,gu2019badnets} to spatially warped or sample-specific invisible perturbations~\citep{li2021issba,nguyen2021wanet}. These attacks transfer effectively to VLMs~\citep{liang2025revisiting}. More recent work has introduced attacks specifically designed for them: TrojVLM injects backdoors during instruction tuning while preserving caption quality on clean images~\citep{lyu2024trojvlm}, and VLOOD demonstrates that backdoors can be implanted using only out-of-distribution auxiliary data~\citep{lyu2025vlood}. Domain-shift studies further indicate that VLM backdoors can remain effective across mismatched training and testing domains~\citep{liang2025revisiting}. These results motivate defenses that suppress backdoor behavior within the model rather than only filtering obvious input triggers.

\noindent\textbf{Backdoor defenses for vision-language models.}
VLM-specific defenses can be broadly categorized into training-time and test-time approaches. Training-time methods either identify and remove poisoned samples before adaptation or regularize the adapted parameters to suppress backdoor triggers. For example, BYE detects suspicious samples in VLM fine-tuning using attention-based signals and clustering~\citep{rong2025bye}, while RobustIT regularizes adapter fine-tuning to reduce trigger effects~\citep{xun2025robustit}. Test-time methods, including PurMM, SRD, and CleanSight, intervene in the input or model computation during inference~\citep{jiang2026purmm,xu2026srd,zhang2026cleansight}. Training-time methods require intervention during adaptation, whereas test-time methods introduce per-query overhead. In contrast, PSR purifies VLM projection weights after training without modifying the adaptation process or adding inference-time computation.

\noindent\textbf{General post-training backdoor defenses.}
Post-training backdoor defenses based on neurons, channels, representations, or weight dynamics provide important technical foundations for PSR. Pruning methods remove backdoor-related components based on various criteria: Fine-Pruning removes neurons dormant on clean data~\citep{liu2018finepruning}, Neural Cleanse detects anomalous triggers and applies pruning~\citep{wang2019neuralcleanse}, and ANP prunes neurons sensitive to adversarial perturbations on clean data~\citep{wu2021anp}. Channel-level methods such as CLP apply Lipschitzness-based criteria~\citep{zheng2022clp}. Other methods use unlearning-based signals to locate backdoor-related components, including RNP and TSBD~\citep{li2023rnp,lin2024tsbd}. Spectral Signatures detects poisoned samples via anomalous representation directions~\citep{tran2018spectral}. However, because these defenses were not specifically designed for VLMs, their direct application may yield suboptimal performance; in contrast, PSR is tailored to the distinctive architectural characteristics of VLMs, making it better suited for backdoor mitigation in this setting.

\section{Preliminaries}
\label{sec:preliminary}

\subsection{Threat Model}
\label{subsec:threat_model}

\noindent\textbf{Victim model.}
We consider a VLM as a composition of a visual encoder, a visual-to-language projection interface, and a language model. Given an image \(\mathbf{x}\) and a text query or instruction \(\mathbf{q}\), the model output is written abstractly as
\begin{gather}
    \mathbf{o} = M_{\psi}\!\left(P_{\phi}\!\left(E_{\mathrm{vis}}(\mathbf{x})\right), \mathbf{q}\right),
\end{gather}
where \(E_{\mathrm{vis}}\) is the visual encoder, \(P_{\phi}\) is the visual-to-language projection interface with parameters \(\phi\), and \(M_{\psi}\) is the language model. During downstream adaptation, only the projection interface is fine-tuned, while the visual encoder and language model remain fixed.

\noindent\textbf{Adversary's objective.}
The adversary aims to implant a backdoor by poisoning the data used to fine-tune the projection interface. We assume that the adversary can control the fine-tuning data and optimization procedure but can modify only the projection parameters. Let \(\mathcal{D}=\mathcal{D}_{\mathrm{clean}}\cup\mathcal{D}_{\mathrm{poison}}\) denote the fine-tuning set, where \(\mathcal{D}_{\mathrm{clean}}\) contains clean image-query-response triples and \(\mathcal{D}_{\mathrm{poison}}\) contains samples with a visual trigger \(\boldsymbol{\tau}\) and an attacker-specified target response \(\mathbf{o}^{\star}\). After fine-tuning on \(\mathcal{D}\), the resulting parameters \(\phi_b\) should preserve normal behavior on a clean image \(\mathbf{x}\) while producing the target response for its triggered version \(\mathbf{x}\oplus\boldsymbol{\tau}\):
\begin{gather}
    f_{\phi_b}(\mathbf{x},\mathbf{q})=\mathbf{o},
    \qquad
    f_{\phi_b}(\mathbf{x}\oplus\boldsymbol{\tau},\mathbf{q})=\mathbf{o}^{\star}.
\end{gather}

\noindent\textbf{Defender's setting.}
The defender receives a fine-tuned VLM suspected of containing a backdoor and aims to purify it. We assume access to the corresponding pretrained base VLM and its projection parameters \(\phi_0\), a small clean calibration set \(\mathcal{C}\), and a restoration-ratio hyperparameter \(r\). The defender need not have the original fine-tuning set or the complete adaptation recipe. The goal is to construct \(\phi_{\mathrm{pur}} = \mathcal{A}(\phi_b, \phi_0, \mathcal{C}; r)\) that reduces backdoor behavior while retaining clean-task utility.
 \subsection{Projection Interface and Channel Representation}
\label{subsec:projection_channel_representation}

PSR targets selected linear layers inside the visual-to-language projection interface \(P_{\phi}\). Let \(\mathcal{L}=\{\ell_1,\ell_2,\ldots,\ell_m\}\) denote these target layers. For layer \(\ell\), its affine output is
\begin{gather}
    \mathbf{u}^{(\ell)} = \mathbf{W}^{(\ell)}\mathbf{v}^{(\ell)} + \mathbf{b}^{(\ell)},
\end{gather}
where \(\mathbf{W}^{(\ell)}\in\mathbb{R}^{d_{\mathrm{out}}^{(\ell)}\times d_{\mathrm{in}}^{(\ell)}}\) and \(\mathbf{b}^{(\ell)}\in\mathbb{R}^{d_{\mathrm{out}}^{(\ell)}}\). Each output channel is indexed by \(i\in\{1,\ldots,d_{\mathrm{out}}^{(\ell)}\}\) and corresponds to the activation \(u_i^{(\ell)}\), weight row \(\mathbf{W}_{i,:}^{(\ell)}\), and bias entry \(b_i^{(\ell)}\). This row-wise correspondence identifies each output channel and its associated parameters for the subsequent method.

\section{Methodology}
\label{sec:method}

\begin{figure}[!t]
    \centering
    \includegraphics[width=\linewidth]{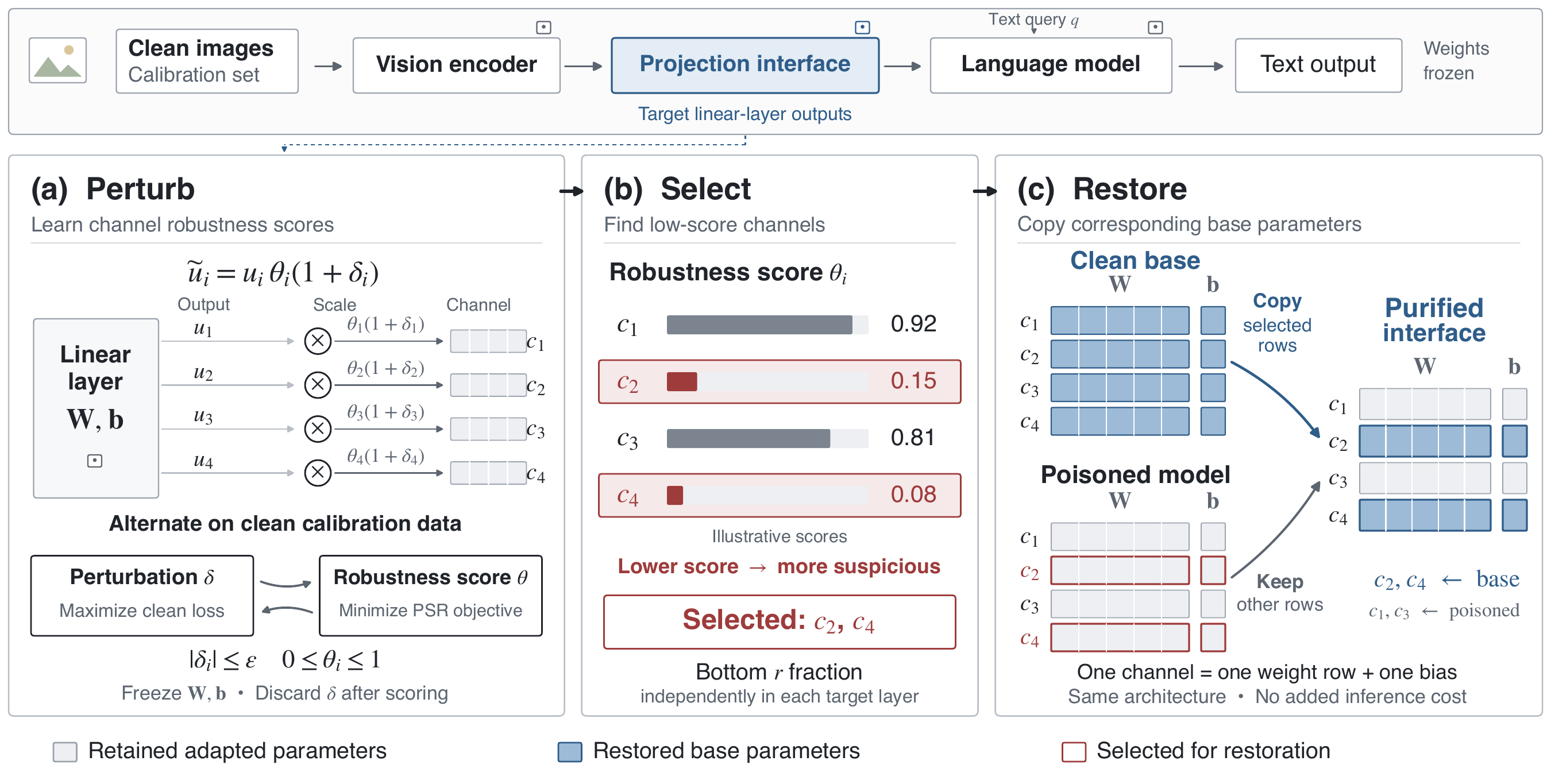}
    \caption{Overview of Perturb-Select-Restore (PSR), a post-training purification framework for VLM projection interfaces.}
    \label{fig:psr_overview}
\end{figure}

\noindent\textbf{Overview.}
PSR is a three-stage post-training purification procedure, illustrated in \autoref{fig:psr_overview}. Starting from a backdoored projector \(\phi_b\), it introduces a channel-wise robustness-score vector \(\boldsymbol{\theta}\) and bounded auxiliary perturbations \(\boldsymbol{\delta}\) at each target projection layer. Using only clean calibration data, PSR alternates between maximizing the clean loss over \(\boldsymbol{\delta}\) to expose sensitivity and minimizing a regularized clean-loss objective over \(\boldsymbol{\theta}\), while keeping the model parameters frozen. Channels that are more vulnerable under this perturbation probe exert stronger pressure for their robustness scores to decrease, so the smallest learned \(\boldsymbol{\theta}\) values identify the channels most likely to contain backdoor-related changes. PSR selects a fixed fraction of these channels independently in each layer, restores their weight rows and bias entries from \(\phi_0\), and leaves all other adapted parameters unchanged. This perturb-select-restore procedure concentrates the repair on fragile channels while preserving task-relevant adaptation elsewhere in the projection interface.

\subsection{Channel-wise Parameterization}
\label{subsec:channel_parameterization}

For each target layer \(\ell\), PSR applies the following channel-wise transformation to the affine output \(\mathbf{u}^{(\ell)}\):
\begingroup
\setlength{\abovedisplayskip}{8pt plus 3pt minus 2pt}
\setlength{\belowdisplayskip}{8pt plus 3pt minus 2pt}
\setlength{\abovedisplayshortskip}{6pt plus 3pt minus 2pt}
\setlength{\belowdisplayshortskip}{6pt plus 3pt minus 2pt}
\begin{gather}
    \label{eq:channel_parameterization}
    \widetilde{\mathbf{u}}^{(\ell)}
    = \mathbf{u}^{(\ell)} \odot
    \left(\boldsymbol{\theta}^{(\ell)} \odot
    \left(\mathbf{1}+\boldsymbol{\delta}^{(\ell)}\right)\right),
\end{gather}
\endgroup

Here, \(\boldsymbol{\theta}^{(\ell)} \in [0,1]^{d_{\mathrm{out}}^{(\ell)}}\) is a learnable channel robustness score vector, and \(\boldsymbol{\delta}^{(\ell)} \in [-\epsilon,\epsilon]^{d_{\mathrm{out}}^{(\ell)}}\) is a bounded auxiliary perturbation vector. The operator \(\odot\) denotes element-wise multiplication with broadcasting along the non-channel dimensions. In the parameterization, \(\boldsymbol{\theta}^{(\ell)}\) acts as a channel-wise retention coefficient: values close to \(1\) preserve the corresponding channels, whereas values close to \(0\) suppress them.

The model parameters remain frozen, and only these channel-wise variables are optimized. The perturbation vector \(\boldsymbol{\delta}^{(\ell)}\) provides bounded adversarial changes to the channel outputs and is used only to learn \(\boldsymbol{\theta}^{(\ell)}\). After optimization, the learned \(\boldsymbol{\theta}^{(\ell)}\) values are used to select channels for restoration, whereas \(\boldsymbol{\delta}^{(\ell)}\) is discarded. We use \(\boldsymbol{\theta}\) and \(\boldsymbol{\delta}\) to denote the collections of these variables across all target layers.

\begin{algorithm}[!t]
\caption{PSR optimization and restoration}
\label{alg:psr_procedure}
\begin{algorithmic}[1]
\small
\Require Poisoned projection parameters $\phi_b$, projection parameters $\phi_0$ of the corresponding pretrained base VLM, clean calibration set $\mathcal{C}$, target layers $\mathcal{L}$, restoration ratio $r$, perturbation bound $\epsilon$, optimization rounds $T_{\mathrm{opt}}$, inner steps $T_{\mathrm{PGD}}$, robustness-score learning rate $\eta_{\theta}$, loss weights $\lambda_{\mathrm{clean}}$ and $\lambda_{\theta}$
\Ensure Purified projection parameters $\phi_{\mathrm{pur}}$
\State Introduce channel-wise variables at the output of each target linear layer in $\mathcal{L}$
\State Initialize each robustness-score vector $\boldsymbol{\theta}^{(\ell)}\gets\mathbf{1}_{d_{\mathrm{out}}^{(\ell)}}$ and keep all model parameters frozen
\For{$s=1,\ldots,T_{\mathrm{opt}}$}
    \State Initialize bounded channel perturbations $\boldsymbol{\delta}^{(\ell)}$ for all $\ell\in\mathcal{L}$
    \For{$t=1,\ldots,T_{\mathrm{PGD}}$}
        \State $\boldsymbol{\delta}\gets\Pi_{[-\epsilon,\epsilon]}\!\left(\boldsymbol{\delta}+\frac{2\epsilon}{T_{\mathrm{PGD}}}\operatorname{sign}\!\left(\nabla_{\boldsymbol{\delta}}\mathcal{L}_{\mathrm{clean}}(\boldsymbol{\theta},\boldsymbol{\delta};\mathcal{C})\right)\right)$
    \EndFor
    \State $\mathbf{g}_{\theta}\gets\nabla_{\boldsymbol{\theta}}\left[\mathcal{L}_{\mathrm{clean}}(\boldsymbol{\theta},\boldsymbol{\delta};\mathcal{C})+\lambda_{\mathrm{clean}}\mathcal{L}_{\mathrm{clean}}(\boldsymbol{\theta},\mathbf{0};\mathcal{C})+\lambda_{\theta}\sum_{\ell\in\mathcal{L}}\left\|\boldsymbol{\theta}^{(\ell)}\right\|_1\right]$
    \State $\boldsymbol{\theta}\gets\Pi_{[0,1]}\!\left(\boldsymbol{\theta}-\eta_{\theta}\operatorname{clip}_{[-1,1]}(\mathbf{g}_{\theta})\right)$
\EndFor
\State Initialize $\phi_{\mathrm{pur}}\gets\phi_b$
\For{each target layer $\ell\in\mathcal{L}$}
    \State Select $S^{(\ell)}\gets\operatorname{TopK}_{\mathrm{smallest}}(\boldsymbol{\theta}^{(\ell)},\lceil r d_{\mathrm{out}}^{(\ell)}\rceil)$
    \For{each selected channel $i\in S^{(\ell)}$}
        \State Restore channel $i$ in $\phi_{\mathrm{pur}}$ from the corresponding channel in $\phi_0$
    \EndFor
\EndFor
\State Save and return $\phi_{\mathrm{pur}}$
\end{algorithmic}
\end{algorithm}

\subsection{Robustness Score Optimization}
\label{subsec:minmax_optimization}

Building on the channel-wise parameterization in \autoref{eq:channel_parameterization}, PSR uses the small clean calibration set \(\mathcal{C}\) as its only data input to learn the channel robustness scores. Let \(\mathcal{L}_{\mathrm{clean}}(\boldsymbol{\theta},\boldsymbol{\delta};\mathcal{C})\) denote the model's clean language-modeling loss under the channel variables. The optimization alternates between an inner maximization over the bounded perturbations \(\boldsymbol{\delta}\) and an outer minimization over the robustness-score vectors \(\boldsymbol{\theta}\).

\noindent\textbf{Inner maximization.}
For fixed robustness scores, the inner problem searches for a bounded channel perturbation that increases the clean loss:
\begingroup
\setlength{\abovedisplayskip}{8pt plus 3pt minus 2pt}
\setlength{\belowdisplayskip}{8pt plus 3pt minus 2pt}
\setlength{\abovedisplayshortskip}{6pt plus 3pt minus 2pt}
\setlength{\belowdisplayshortskip}{6pt plus 3pt minus 2pt}
\begin{gather}
    \label{eq:inner_maximization}
    \boldsymbol{\delta}^{\star}
    = \mathop{\arg\max}_{\|\boldsymbol{\delta}\|_{\infty}\leq\epsilon}
    \mathcal{L}_{\mathrm{clean}}(\boldsymbol{\theta},\boldsymbol{\delta};\mathcal{C}).
\end{gather}
\endgroup
This problem is approximated by projected sign-gradient ascent with step size \(2\epsilon/T_{\mathrm{PGD}}\), as specified in \autoref{alg:psr_procedure}. After each update, every \(\boldsymbol{\delta}^{(\ell)}\) is projected elementwise onto the box \([-\epsilon,\epsilon]^{d_{\mathrm{out}}^{(\ell)}}\).

\noindent\textbf{Outer minimization.}
With the current adversarial perturbation fixed, the robustness scores are updated using the clean-data loss under channel perturbations, the clean loss with the auxiliary perturbation set to zero, and an \(\ell_1\) score penalty:
\begingroup
\setlength{\abovedisplayskip}{8pt plus 3pt minus 2pt}
\setlength{\belowdisplayskip}{8pt plus 3pt minus 2pt}
\setlength{\abovedisplayshortskip}{6pt plus 3pt minus 2pt}
\setlength{\belowdisplayshortskip}{6pt plus 3pt minus 2pt}
\begin{gather}
    \min_{\{\boldsymbol{\theta}^{(\ell)}\}_{\ell\in\mathcal{L}}}
    \mathcal{L}_{\mathrm{clean}}(\boldsymbol{\theta},\boldsymbol{\delta}^{\star};\mathcal{C})
    + \lambda_{\mathrm{clean}}
    \mathcal{L}_{\mathrm{clean}}(\boldsymbol{\theta},\mathbf{0};\mathcal{C})
    + \lambda_{\theta}
    \sum_{\ell\in\mathcal{L}}
    \left\|\boldsymbol{\theta}^{(\ell)}\right\|_1,
    \quad
    \text{s.t. }\boldsymbol{\theta}^{(\ell)}\in[0,1]^{d_{\mathrm{out}}^{(\ell)}}.
\end{gather}
\endgroup
In each round, the implementation first performs \(T_{\mathrm{PGD}}\) projected sign-gradient ascent steps on \(\boldsymbol{\delta}\), then updates \(\boldsymbol{\theta}\) using the outer-objective gradient clipped elementwise to \([-1,1]\), and projects each robustness-score vector back to its box constraint \([0,1]^{d_{\mathrm{out}}^{(\ell)}}\).

The learned \(\boldsymbol{\theta}\) values serve as channel-wise retention coefficients and provide the signal for subsequent purification. Under the inner maximization, channels whose retained outputs are more sensitive to bounded perturbations create greater pressure for the outer minimization to reduce their scores, while the unperturbed clean-loss term discourages indiscriminate suppression of all channels. Consequently, a lower learned \(\theta_i^{(\ell)}\) indicates that channel \(i\) is more fragile under the clean-data perturbation probe and is treated as more suspicious. We therefore select low-score channels for restoration.

\subsection{Channel Selection and Parameter Restoration}
\label{subsec:base_guided_restoration}

Once robustness-score optimization is complete, PSR performs restoration independently in each target layer. Given the restoration ratio \(r\), it selects
\(k^{(\ell)}=\left\lceil r\,d_{\mathrm{out}}^{(\ell)}\right\rceil\) channels from layer \(\ell\). Let
\(S^{(\ell)}=\operatorname{TopK}_{\mathrm{smallest}}\!\left(\boldsymbol{\theta}^{(\ell)},k^{(\ell)}\right)\)
denote the indices of the selected channels. For any projection parameters \(\phi\), let \(\mathbf{W}^{(\ell)}(\phi)\) and \(\mathbf{b}^{(\ell)}(\phi)\) denote the weight matrix and bias vector of layer \(\ell\), respectively. For each selected channel \(i\in S^{(\ell)}\), PSR restores its parameters from the corresponding pretrained base VLM:
\(\mathbf{W}_{i,:}^{(\ell)}(\phi_{\mathrm{pur}})=\mathbf{W}_{i,:}^{(\ell)}(\phi_0)\) and \(b_i^{(\ell)}(\phi_{\mathrm{pur}})=b_i^{(\ell)}(\phi_0)\). For each unselected channel \(i\notin S^{(\ell)}\), it leaves the adapted parameters unchanged, so \(\mathbf{W}_{i,:}^{(\ell)}(\phi_{\mathrm{pur}})=\mathbf{W}_{i,:}^{(\ell)}(\phi_b)\) and \(b_i^{(\ell)}(\phi_{\mathrm{pur}})=b_i^{(\ell)}(\phi_b)\).

\autoref{alg:psr_procedure} summarizes the complete PSR procedure, including robustness-score optimization, per-layer channel selection, and restoration from \(\phi_0\). Here, \(\Pi\) denotes projection onto the indicated feasible interval, \(\operatorname{clip}\) denotes elementwise clipping, \(T_{\mathrm{opt}}\) is the number of optimization rounds, \(T_{\mathrm{PGD}}\) is the number of inner ascent steps, and \(\eta_{\theta}\) is the robustness-score learning rate.
\section{Experiments}
\label{sec:experiments}

% Declare the two main-result tables early so they can be placed at the top of the
% page where the experimental setup continues.
\begin{table}[!htbp]
\centering
\caption{Results on image captioning (COCO). ASR (\%, $\downarrow$) and clean utility (CU; CIDEr on clean input, $\uparrow$) are reported across attacks and defenses.}
\label{tab:main_results_coco}
\scriptsize
\resizebox{\textwidth}{!}{%
\setlength{\tabcolsep}{3pt}%
\begin{tabular}{llcccccccccccc}
\toprule
 & & \multicolumn{2}{c}{BadNet} & \multicolumn{2}{c}{Blended} & \multicolumn{2}{c}{ISSBA} & \multicolumn{2}{c}{TrojVLM} & \multicolumn{2}{c}{VLOOD} & \multicolumn{2}{c}{WaNet} \\
Model & Method & ASR & CU & ASR & CU & ASR & CU & ASR & CU & ASR & CU & ASR & CU \\
\midrule
\multirow{6}{*}{LLaVA-1.5} & No defense & 99.38 & 120.70 & 95.42 & 124.17 & 97.24 & 123.69 & 93.46 & 117.77 & 98.70 & 120.88 & 92.36 & 123.70 \\
 & Clean FT & 99.36 & 120.93 & \textbf{0.00} & 123.98 & 1.82 & 123.89 & 93.46 & 117.88 & 98.68 & 121.01 & \textbf{0.00} & 123.82 \\
 & Random & 0.66 & 121.96 & 1.82 & 123.28 & 20.62 & 124.82 & 3.40 & 124.04 & 34.08 & 124.10 & 16.92 & 124.66 \\
 & FP & 0.76 & 121.92 & 5.20 & 122.58 & 17.62 & 124.96 & 3.20 & 122.89 & 33.96 & 123.35 & 25.12 & 123.40 \\
 & CLP & \textbf{0.08} & 122.08 & 1.88 & 123.95 & 3.92 & 125.07 & \textbf{1.24} & 123.87 & 9.90 & 124.24 & 1.26 & 124.96 \\
\rowcolor{gray!12} & PSR & 0.84 & 120.55 & 1.54 & 120.67 & \textbf{0.00} & 122.93 & 1.60 & 121.04 & \textbf{0.28} & 121.84 & 0.56 & 121.70 \\
\midrule
\multirow{6}{*}{Qwen3-VL} & No defense & 99.32 & 128.38 & 91.70 & 127.74 & 91.32 & 128.13 & 97.44 & 126.05 & 91.48 & 127.73 & 95.62 & 121.71 \\
 & Clean FT & 99.32 & 128.23 & \textbf{0.02} & 127.96 & \textbf{0.00} & 127.92 & 97.30 & 125.56 & 91.76 & 127.39 & 0.06 & 121.15 \\
 & Random & 53.52 & 128.32 & 16.40 & 127.07 & 38.32 & 128.48 & 67.82 & 127.25 & 22.38 & 128.13 & 47.58 & 127.45 \\
 & FP & 57.96 & 129.11 & 0.74 & 127.47 & 39.02 & 129.02 & 40.00 & 127.38 & 50.34 & 128.33 & 43.94 & 126.40 \\
 & CLP & 36.68 & 128.77 & 17.88 & 127.45 & 23.10 & 128.20 & 15.82 & 128.62 & 19.92 & 128.14 & 45.24 & 127.44 \\
\rowcolor{gray!12} & PSR & \textbf{0.90} & 124.42 & 1.70 & 123.03 & \textbf{0.00} & 126.00 & \textbf{0.32} & 125.37 & \textbf{0.00} & 127.30 & \textbf{0.04} & 126.09 \\
\bottomrule
\end{tabular}}
\vspace{8pt}
\centering
\caption{Results on visual question answering (VQAv2). ASR (\%, $\downarrow$) and clean utility (CU; VQA accuracy on clean input, \%, $\uparrow$) are reported across attacks and defenses.}
\label{tab:main_results_vqa}
\scriptsize
\resizebox{\textwidth}{!}{%
\setlength{\tabcolsep}{3pt}%
\begin{tabular}{llcccccccccccc}
\toprule
 & & \multicolumn{2}{c}{BadNet} & \multicolumn{2}{c}{Blended} & \multicolumn{2}{c}{ISSBA} & \multicolumn{2}{c}{TrojVLM} & \multicolumn{2}{c}{VLOOD} & \multicolumn{2}{c}{WaNet} \\
Model & Method & ASR & CU & ASR & CU & ASR & CU & ASR & CU & ASR & CU & ASR & CU \\
\midrule
\multirow{6}{*}{LLaVA-1.5} & No defense & 99.86 & 74.84 & 99.76 & 74.07 & 99.32 & 73.81 & 99.90 & 74.14 & 94.18 & 41.60 & 98.34 & 74.47 \\
 & Clean FT & 99.86 & 73.84 & 3.52 & 72.87 & 0.80 & 72.96 & 99.96 & 72.44 & 96.78 & 9.32 & 36.52 & 73.41 \\
 & Random & 2.94 & 73.19 & 9.28 & 74.90 & 5.38 & 74.84 & 0.24 & 73.35 & 1.24 & 24.91 & 52.96 & 75.06 \\
 & FP & 4.78 & 72.64 & 11.02 & 74.89 & 5.36 & 75.17 & 3.74 & 73.90 & \textbf{0.32} & 40.19 & 65.28 & 75.14 \\
 & CLP & \textbf{0.08} & 74.26 & 3.54 & 74.65 & \textbf{0.00} & 75.01 & \textbf{0.02} & 73.63 & 0.76 & 22.63 & 8.92 & 75.07 \\
\rowcolor{gray!12} & PSR & 0.12 & 72.85 & \textbf{0.58} & 75.11 & 0.12 & 75.28 & 0.14 & 74.35 & 1.40 & 26.57 & \textbf{0.08} & 75.45 \\
\midrule
\multirow{6}{*}{Qwen3-VL} & No defense & 96.34 & 80.96 & 90.64 & 79.50 & 93.80 & 80.25 & 92.78 & 80.74 & 94.70 & 49.00 & 93.24 & 80.43 \\
 & Clean FT & 96.22 & 81.00 & 89.68 & 79.60 & 92.76 & 80.45 & 92.70 & 80.68 & 93.26 & 78.50 & \textbf{0.00} & 0.20 \\
 & Random & 6.26 & 79.77 & 10.96 & 79.67 & 13.12 & 80.13 & 4.54 & 79.57 & 4.02 & 21.58 & 9.26 & 80.00 \\
 & FP & 1.04 & 79.49 & 2.24 & 79.42 & 3.48 & 80.02 & 0.90 & 79.40 & 2.26 & 72.58 & 5.68 & 80.05 \\
 & CLP & 15.42 & 79.74 & 18.70 & 79.74 & 15.88 & 80.19 & 8.76 & 79.67 & 8.42 & 23.98 & 12.92 & 80.03 \\
\rowcolor{gray!12} & PSR & \textbf{0.62} & 79.66 & \textbf{0.00} & 79.61 & \textbf{0.00} & 80.10 & \textbf{0.50} & 79.73 & \textbf{0.10} & 53.66 & \textbf{0.00} & 80.00 \\
\bottomrule
 \end{tabular}}
\end{table}

\subsection{Experimental Setup}
\label{subsec:experimental_setup}

\noindent\textbf{Victim models and benchmarks.}
We conduct experiments on two VLM architectures and two multimodal benchmarks. The evaluated models are Qwen3-VL-8B-Instruct~\citep{qwenteam2025qwen3vl}, which uses a visual merger together with DeepStack modules as its visual-to-language projection interface, and LLaVA-1.5-7B~\citep{liu2024llava}, which uses a two-layer MLP projector. The evaluated tasks are image captioning on MS-COCO~\citep{lin2014coco,chen2015coco} and visual question answering on VQAv2~\citep{goyal2017vqav2}; each model is evaluated on both tasks. In all settings, the backdoor is implanted by parameter-efficient adaptation of the visual-to-language projection interface while the visual encoder and language model remain fixed. Appendix~\hyperref[subsec:appendix_models]{\ref*{subsec:appendix_models}} gives the model and target-layer details, while Appendix~\hyperref[subsec:appendix_datasets]{\ref*{subsec:appendix_datasets}} describes the datasets and evaluation protocols.

\noindent\textbf{Backdoor setup.}
Following prior VLM backdoor evaluations, we consider six attack constructions: BadNet~\citep{gu2019badnets}, Blended~\citep{chen2017targeted}, ISSBA~\citep{li2021issba}, WaNet~\citep{nguyen2021wanet}, TrojVLM~\citep{lyu2024trojvlm}, and VLOOD~\citep{lyu2025vlood}. All attacks use the target response ``you have been hacked lol.'' Appendix~\hyperref[subsec:appendix_checkpoints]{\ref*{subsec:appendix_checkpoints}} describes the trigger mechanisms and target-injection strategies for these attacks.

\noindent\textbf{Defense baselines.}
We compare PSR with Clean Fine-tuning (Clean FT), a conventional clean-data repair baseline, and three post-training channel-selection baselines: Random, Fine-Pruning-style (FP)~\citep{liu2018finepruning}, and CLP-style (CLP)~\citep{zheng2022clp}. Random selects channels uniformly at random within each target layer. We focus on these methods because they can be evaluated under PSR's defender setting, starting from the same fine-tuned poisoned model and performing model-level repair before deployment. For a controlled comparison, the channel-selection baselines rank the same target channels and restore the same number per layer from \(\phi_0\). We also report the undefended poisoned model as a no-defense reference. Detailed scoring rules and comparison protocols are provided in Appendix~\hyperref[subsec:appendix_defense_protocol]{\ref*{subsec:appendix_defense_protocol}}.

\noindent\textbf{Evaluation metrics.}
We report attack success rate (ASR, $\downarrow$), which measures the proportion of triggered inputs whose generated output contains the target response, and clean utility (CU, $\uparrow$), measured by CIDEr~\citep{vedantam2015cider} for image captioning and official VQA accuracy for visual question answering. Lower ASR indicates more effective backdoor mitigation, while higher CU indicates better preservation of clean-task performance.

\noindent\textbf{Implementation details.}
PSR learns channel robustness scores from a small clean calibration set of 256 examples, disjoint from validation. The common configuration uses a perturbation bound $\epsilon=0.016$, two projected-gradient ascent steps, robustness-score learning rate $\eta_\theta=0.06$, clean-loss weight $\lambda_{\mathrm{clean}}=0.5$, and robustness-score regularization $\lambda_\theta=0.006$. Channels are ranked independently within each target layer, and the weight rows and bias entries of the selected channels are restored to their corresponding values in \(\phi_0\) according to the restoration-ratio hyperparameter $r$. Full optimization hyperparameters are given in Appendix~\hyperref[subsec:appendix_hyperparameters]{\ref*{subsec:appendix_hyperparameters}}.

\subsection{Main Results}
\label{subsec:main_results}

% The two main-result tables are declared at the start of this section.

Tables~\hyperref[tab:main_results_coco]{\ref*{tab:main_results_coco}} and~\hyperref[tab:main_results_vqa]{\ref*{tab:main_results_vqa}} report results across two VLM architectures, two multimodal tasks, and six attack constructions. PSR reduces ASR to at most 1.70\% in every evaluated setting, showing that the perturbation-based channel scores consistently identify projection channels associated with the backdoor.

\noindent\textbf{Existing defenses are not consistently reliable.} Clean FT removes some attacks but leaves the backdoor largely intact in many cases: on COCO, it leaves ASR above 91\% for TrojVLM and VLOOD on both models, and on LLaVA VQAv2 it leaves 99.86\% on BadNet and 99.96\% on TrojVLM. Its occasional successes can also coincide with severe utility collapse, such as Qwen3-VL on VQAv2 under WaNet, where ASR falls to 0.00\% but CU drops from 80.43\% to 0.20\%. Channel-selection baselines are more effective on individual attacks but vary substantially across models and datasets. For example, CLP leaves 45.24\% ASR on Qwen3-VL COCO under WaNet and 18.70\% on Qwen3-VL VQAv2 under Blended, while Random and FP leave 67.82\% and 40.00\% on TrojVLM COCO, respectively. No single baseline therefore provides the reliable cross-setting suppression achieved by PSR.

\noindent\textbf{PSR preserves clean utility while suppressing the backdoor.} On LLaVA COCO, PSR reduces ASR to at most 1.60\% across all attacks, including 0.00\% on ISSBA, while retaining CIDEr values between 120.55 and 122.93. On Qwen3-VL COCO, it achieves 0.00\% ASR on ISSBA and VLOOD and 0.04\% on WaNet, with CIDEr remaining between 123.03 and 127.30. The same pattern holds for VQAv2: PSR keeps ASR below 1.40\% on LLaVA and below 0.62\% on Qwen3-VL. Clean accuracy remains close to that of the undefended models in most settings, with CU around 73--75\% for LLaVA and 80\% for Qwen3-VL. Overall, PSR is effective across both the compact MLP projector and the distributed merger interface, demonstrating its consistency across different projection architectures.

\subsection{Ablation Studies}
\label{subsec:ablation_studies}

All ablation experiments use Qwen3-VL-8B-Instruct on COCO under four attacks: BadNet, TrojVLM, Blended, and WaNet. Unless noted, we use the standard PSR configuration.

\begin{figure}[!tbp]
    \centering
    \includegraphics[width=\linewidth]{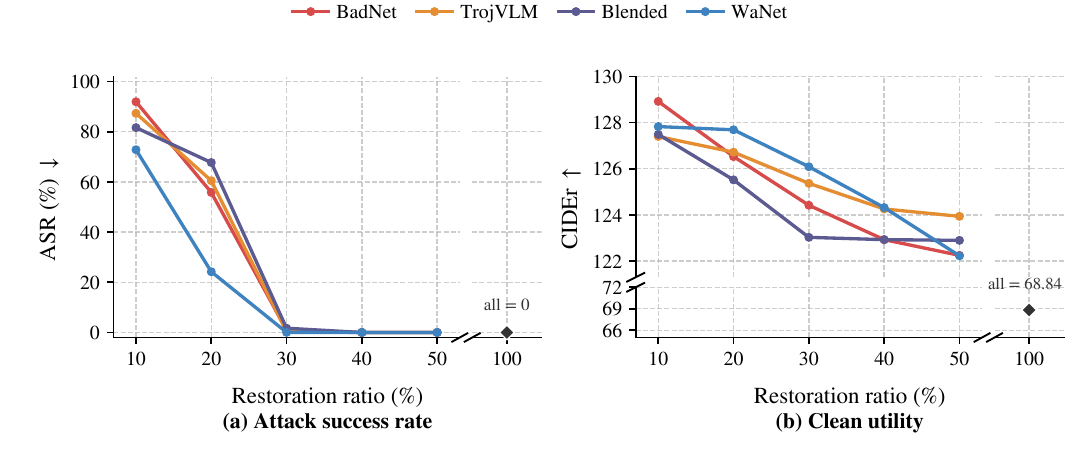}
    \caption{Ablation of the PSR restoration ratio on Qwen3-VL-8B COCO attack settings. (a) ASR ($\downarrow$) on triggered images; (b) clean utility (CU; CIDEr, $\uparrow$) on clean images. Each curve reuses the same optimized robustness scores and restores selected parameters from \(\phi_0\). Diamonds denote the shared result of all four attacks at 100\% restoration.}
    \label{fig:psr_qwen3_ratio_ablation}
\end{figure}

\textbf{Restoration ratio.} \autoref{fig:psr_qwen3_ratio_ablation} varies the restoration ratio from 10\% to 50\% with fixed robustness scores. Across attacks, increasing the restoration ratio consistently lowers ASR, although the rate of reduction varies by attack, while clean CIDEr declines gradually. This result characterizes the security--utility trade-off controlled by the restoration-ratio hyperparameter. At $r=100\%$ all target channels are restored, so the projection reverts to the pretrained base VLM: ASR is 0.00\% for all four attacks, but clean CIDEr drops to 68.84 versus 123.03--126.09 at $r=30\%$.

\begin{table}[!htbp]
\centering
\caption{Discrete ablations on Qwen3-VL COCO across four attacks. ASR/CU are reported for first-group, second-group, and joint restoration, plus restoration from \(\phi_0\) versus zero.}
\label{tab:psr_layer_operation_ablation}
\footnotesize
\setlength{\tabcolsep}{3pt}
\begin{tabular*}{\textwidth}{@{\extracolsep{\fill}}lccccccc@{}}
\toprule
 & \multicolumn{3}{c}{Layer coverage: ASR (\%, $\downarrow$)} & \multicolumn{2}{c}{CU ($\uparrow$)} & \multicolumn{2}{c}{ASR (\%, $\downarrow$)} \\
\cmidrule(lr){2-4}\cmidrule(lr){5-6}\cmidrule(lr){7-8}
Attack & First group & Second group & Both groups & From \(\phi_0\) & To zero & From \(\phi_0\) & To zero \\
\midrule
BadNet & 97.52 & 67.02 & 0.90 & 124.42 & 107.89 & 0.90 & 0.00 \\
TrojVLM & 89.28 & 64.08 & 0.32 & 125.37 & 115.97 & 0.32 & 0.00 \\
Blended & 82.74 & 55.84 & 1.70 & 123.03 & 97.37 & 1.70 & 0.00 \\
WaNet & 84.10 & 31.30 & 0.04 & 126.09 & 118.28 & 0.04 & 0.00 \\
\bottomrule
\end{tabular*}
\end{table}

\textbf{Layer coverage.} \autoref{tab:psr_layer_operation_ablation} compares repairing the first and second MLP layer groups across the merger modules, separately and jointly, under the corresponding settings. When only one MLP layer group is restored, channel robustness scores are learned only for that group across the main merger and the three DeepStack mergers. Restoring only the first group leaves 82.74--97.52\% ASR, and restoring only the second leaves 31.30--67.02\% ASR; jointly restoring both groups reduces ASR to 0.04--1.70\% across all four attacks. The result indicates that the backdoor signal is distributed across the merger interface rather than concentrated in a single layer.

\textbf{Restoration choice.} Using the same selected channels, restoring their weights and biases from \(\phi_0\) yields 0.04--1.70\% ASR and clean CIDEr scores of 123.03--126.09 across the four attacks. Setting the same parameters to zero also suppresses ASR to 0.00\%, but reduces clean CIDEr to 97.37--118.28; for example, on BadNet, restoration from \(\phi_0\) gives CIDEr 124.42 versus 107.89 when setting the parameters to zero. These results support restoration from \(\phi_0\) as the default operation because it preserves substantially more clean utility while achieving comparable attack removal.

\FloatBarrier

\subsection{Robustness to Adaptive Attacks}
\label{subsec:adaptive_attacks}

We finally evaluate PSR against adaptive adversaries who know the defense and adjust attack training to evade it. We extend the adversary in \autoref{subsec:threat_model} with full knowledge of PSR, including its channel-wise parameterization, the bounded perturbation probe, and the base-guided restoration from \(\phi_0\); the adversary controls the fine-tuning data and optimization procedure and may embed a simulation of PSR into attack training. Starting from the standard attack objective, we consider three adaptive strategies of increasing strength.

\noindent\textbf{Perturbation-aware training (PA).} The adversary replays PSR's perturbation probe during attack training: bounded channel perturbations are optimized to increase the clean loss, and the attack objective is augmented with a clean-data term computed under these perturbations, so that the poisoned projector is trained to stay insensitive to the probe that PSR relies on for channel selection.

\noindent\textbf{Persistence-augmented training (PA+).} Building on PA, the adversary additionally requires poisoned samples to keep producing the target response under the same perturbations, directly countering the fragility signal exposed by the probe.

\noindent\textbf{Restoration-aware training (RA).} The strongest adversary simulates the complete PSR procedure during training: it periodically learns channel robustness scores, mimics restoring the lowest-scoring channels toward the base parameters, and optimizes both the clean and attack objectives under this simulated restoration, so that the implanted backdoor is explicitly trained to survive the actual defense.

Experiments are conducted on Qwen3-VL-8B-Instruct on COCO under BadNet, TrojVLM, WaNet, and Blended. The standard-attack reference results are taken from \autoref{tab:main_results_coco}; the three adaptive strategies use the same underlying experimental configuration. All models are purified by PSR under the standard configuration in \autoref{subsec:experimental_setup}. \autoref{tab:adaptive_attacks} reports ASR and CU before and after purification, evaluated on 5{,}000 clean images and their triggered counterparts from COCO.

\begin{table}[!htbp]
\centering
\caption{PSR against adaptive attacks on Qwen3-VL COCO. Each attack is trained with the standard objective (Std.) and three defense-aware strategies of increasing strength (PA, PA+, RA); all models are purified by the same PSR configuration. Each cell reports the metric \emph{after} PSR purification, with the value before purification shown in gray parentheses (ASR in \%; CU in CIDEr).}
\label{tab:adaptive_attacks}
\footnotesize
\newcommand{\abf}[2]{#1~{\color{gray}\scriptsize(#2)}}
\begin{tabular*}{\textwidth}{@{\extracolsep{\fill}}llcccc@{}}
\toprule
Attack & Metric & Std. & PA & PA+ & RA \\
\midrule
\multirow{2}{*}{BadNet}
 & ASR ($\downarrow$) & \abf{0.90}{99.32} & \abf{0.00}{97.78} & \abf{2.08}{99.76} & \abf{0.00}{99.92} \\[1pt]
 & CU ($\uparrow$) & \abf{124.42}{128.38} & \abf{127.96}{123.40} & \abf{126.37}{122.85} & \abf{125.18}{125.55} \\
\midrule
\multirow{2}{*}{TrojVLM}
 & ASR ($\downarrow$) & \abf{0.32}{97.44} & \abf{0.00}{98.22} & \abf{0.00}{99.66} & \abf{0.00}{99.14} \\[1pt]
 & CU ($\uparrow$) & \abf{125.37}{126.05} & \abf{125.55}{123.27} & \abf{124.90}{122.47} & \abf{127.93}{123.90} \\
\midrule
\multirow{2}{*}{WaNet}
 & ASR ($\downarrow$) & \abf{0.04}{95.62} & \abf{0.00}{78.62} & \abf{0.78}{97.22} & \abf{0.00}{90.98} \\[1pt]
 & CU ($\uparrow$) & \abf{126.09}{121.71} & \abf{124.70}{115.76} & \abf{125.74}{116.24} & \abf{126.72}{123.08} \\
\midrule
\multirow{2}{*}{Blended}
 & ASR ($\downarrow$) & \abf{1.70}{91.70} & \abf{0.00}{86.76} & \abf{0.96}{94.46} & \abf{2.94}{93.54} \\[1pt]
 & CU ($\uparrow$) & \abf{123.03}{127.74} & \abf{125.93}{122.50} & \abf{126.32}{122.28} & \abf{127.58}{124.62} \\
\bottomrule
\end{tabular*}
\end{table}

\noindent\textbf{PSR remains effective under defense-aware training.} Across these adaptive settings, PSR reduces ASR to near zero in most settings. These results show that training the backdoor to withstand PSR’s perturbation probe does not necessarily make it robust to the subsequent purification step.

% Slightly enlarge this page so the short Conclusion fits on it instead of
% being pushed to a nearly empty page.
\enlargethispage{10\baselineskip}

\section{Conclusion}
\label{sec:conclusion}

We proposed PSR, a post-training purification method for repairing backdoored VLM visual-to-language projection interfaces. Our analysis identifies projection fragility: backdoored models exhibit disproportionately larger clean-loss increases under bounded projection perturbations, providing a clean-data signal for channel selection. PSR uses this signal to learn robustness scores and directly restores selected projection rows and biases to their corresponding values in \(\phi_0\), producing a purified projection interface without intervention during adaptation or at inference time. Across two VLM architectures, two multimodal tasks, and six attack types, PSR substantially suppresses the attack success rate while generally retaining clean-task utility. These results support clean-data robustness scoring and restoration to pretrained parameter values as a practical direction for mitigating projection-interface backdoors in VLMs under the evaluated threat model.

\clearpage
\bibliography{arxiv_paper}

@inproceedings{liu2024llava,
  title = {Improved Baselines with Visual Instruction Tuning},
  author = {Liu, Haotian and Li, Chunyuan and Li, Yuheng and Lee, Yong Jae},
  booktitle = {Proceedings of the IEEE/CVF Conference on Computer Vision and Pattern Recognition},
  year = {2024}
}

@inproceedings{dai2023instructblip,
  title = {{InstructBLIP}: Towards General-purpose Vision-Language Models with Instruction Tuning},
  author = {Dai, Wenliang and Li, Junnan and Li, Dongxu and Tiong, Anthony Meng Huat and Zhao, Junqi and Wang, Weisheng and Li, Boyang and Fung, Pascale and Hoi, Steven C. H.},
  booktitle = {Advances in Neural Information Processing Systems},
  year = {2023}
}

@inproceedings{zhu2024minigpt4,
  title = {{MiniGPT-4}: Enhancing Vision-Language Understanding with Advanced Large Language Models},
  author = {Zhu, Deyao and Chen, Jun and Shen, Xiaoqian and Li, Xiang and Elhoseiny, Mohamed},
  booktitle = {International Conference on Learning Representations},
  year = {2024}
}

@article{qwenteam2025qwen3vl,
  title = {{Qwen3-VL} Technical Report},
  author = {Bai, Shuai and Cai, Yuxuan and Chen, Ruizhe and Chen, Keqin and Chen, Xionghui and Cheng, Zesen and Deng, Lianghao and Ding, Wei and Gao, Chang and Ge, Chunjiang and Ge, Wenbin and Guo, Zhifang and Huang, Qidong and Huang, Jie and Huang, Fei and Hui, Binyuan and Jiang, Shutong and Li, Zhaohai and Li, Mingsheng and Li, Mei and Li, Kaixin and Lin, Zicheng and Lin, Junyang and Liu, Xuejing and Liu, Jiawei and Liu, Chenglong and Liu, Yang and Liu, Dayiheng and Liu, Shixuan and Lu, Dunjie and Luo, Ruilin and Lv, Chenxu and Men, Rui and Meng, Lingchen and Ren, Xuancheng and Ren, Xingzhang and Song, Sibo and Sun, Yuchong and Tang, Jun and Tu, Jianhong and Wan, Jianqiang and Wang, Peng and Wang, Pengfei and Wang, Qiuyue and Wang, Yuxuan and Xie, Tianbao and Xu, Yiheng and Xu, Haiyang and Xu, Jin and Yang, Zhibo and Yang, Mingkun and Yang, Jianxin and Yang, An and Yu, Bowen and Zhang, Fei and Zhang, Hang and Zhang, Xi and Zheng, Bo and Zhong, Humen and Zhou, Jingren and Zhou, Fan and Zhou, Jing and Zhu, Yuanzhi and Zhu, Ke},
  journal = {arXiv preprint arXiv:2511.21631},
  year = {2025}
}

@article{gu2019badnets,
  title = {{BadNets}: Evaluating Backdooring Attacks on Deep Neural Networks},
  author = {Gu, Tianyu and Liu, Kang and Dolan-Gavitt, Brendan and Garg, Siddharth},
  journal = {IEEE Access},
  volume = {7},
  pages = {47230--47244},
  year = {2019}
}

@article{chen2017targeted,
  title = {Targeted Backdoor Attacks on Deep Learning Systems Using Data Poisoning},
  author = {Chen, Xinyun and Liu, Chang and Li, Bo and Lu, Kimberly and Song, Dawn},
  journal = {arXiv preprint arXiv:1712.05526},
  year = {2017}
}

@inproceedings{nguyen2021wanet,
  title = {{WaNet} - Imperceptible Warping-based Backdoor Attack},
  author = {Nguyen, Tuan Anh and Tran, Anh Tuan},
  booktitle = {International Conference on Learning Representations},
  year = {2021}
}

@inproceedings{li2021issba,
  title = {Invisible Backdoor Attack with Sample-Specific Triggers},
  author = {Li, Yuezun and Li, Yiming and Wu, Baoyuan and Li, Longkang and He, Ran and Lyu, Siwei},
  booktitle = {Proceedings of the IEEE/CVF International Conference on Computer Vision},
  year = {2021}
}

@inproceedings{lyu2024trojvlm,
  title = {{TrojVLM}: Backdoor Attack Against Vision Language Models},
  author = {Lyu, Weimin and Pang, Lu and Ma, Tengfei and Ling, Haibin and Chen, Chao},
  booktitle = {European Conference on Computer Vision},
  year = {2024}
}

@inproceedings{lyu2025vlood,
  title = {Backdooring Vision-Language Models with Out-of-Distribution Data},
  author = {Lyu, Weimin and Yao, Jiachen and Gupta, Saumya and Pang, Lu and Sun, Tao and Yi, Lingjie and Hu, Lijie and Ling, Haibin and Chen, Chao},
  booktitle = {International Conference on Learning Representations},
  year = {2025}
}

@inproceedings{liang2025revisiting,
  title = {Revisiting Backdoor Attacks against Large Vision-Language Models from Domain Shift},
  author = {Liang, Siyuan and Liang, Jiawei and Pang, Tianyu and Du, Chao and Liu, Aishan and Zhu, Mingli and Cao, Xiaochun and Tao, Dacheng},
  booktitle = {Proceedings of the IEEE/CVF Conference on Computer Vision and Pattern Recognition},
  year = {2025}
}

@inproceedings{liu2018finepruning,
  title = {{Fine-Pruning}: Defending Against Backdooring Attacks on Deep Neural Networks},
  author = {Liu, Kang and Dolan-Gavitt, Brendan and Garg, Siddharth},
  booktitle = {Research in Attacks, Intrusions, and Defenses},
  year = {2018}
}

@inproceedings{wu2021anp,
  title = {Adversarial Neuron Pruning Purifies Backdoored Deep Models},
  author = {Wu, Dongxian and Wang, Yisen},
  booktitle = {Advances in Neural Information Processing Systems},
  year = {2021}
}

@inproceedings{zheng2022clp,
  title = {Data-Free Backdoor Removal Based on Channel {Lipschitzness}},
  author = {Zheng, Runkai and Tang, Rongjun and Li, Jianze and Liu, Li},
  booktitle = {European Conference on Computer Vision},
  year = {2022}
}

@inproceedings{wang2019neuralcleanse,
  title = {{Neural Cleanse}: Identifying and Mitigating Backdoor Attacks in Neural Networks},
  author = {Wang, Bolun and Yao, Yuanshun and Shan, Shawn and Li, Huiying and Viswanath, Bimal and Zheng, Haitao and Zhao, Ben Y.},
  booktitle = {IEEE Symposium on Security and Privacy},
  year = {2019}
}

@inproceedings{jiang2026purmm,
  title = {{PurMM}: Attention-Guided Test-Time Backdoor Purification in Multimodal Large Language Models},
  author = {Jiang, Wenzheng and Liang, Ke and Rong, Xuankun and Zhou, Jingxuan and Zhong, Zhengyi and Wan, Guancheng and Wang, Ji},
  booktitle = {Proceedings of the AAAI Conference on Artificial Intelligence},
  year = {2026}
}

@inproceedings{rong2025bye,
  title = {Backdoor Cleaning without External Guidance in {MLLM} Fine-tuning},
  author = {Rong, Xuankun and Huang, Wenke and Liang, Jian and Bi, Jinhe and Xiao, Xun and Li, Yiming and Du, Bo and Ye, Mang},
  booktitle = {Advances in Neural Information Processing Systems},
  year = {2025}
}

@inproceedings{xu2026srd,
  title = {{SRD}: Reinforcement-Learned Semantic Perturbation for Backdoor Defense in {VLMs}},
  author = {Xu, Shuhan and Liang, Siyuan and Zheng, Hongling and Liu, Aishan and Wang, Xinbiao and Luo, Yong and Lin, Fu and Rutkowski, Leszek and Tao, Dacheng},
  booktitle = {Proceedings of the AAAI Conference on Artificial Intelligence},
  year = {2026}
}

@inproceedings{zhang2026cleansight,
  title = {Test-Time Attention Purification for Backdoored Large Vision Language Models},
  author = {Zhang, Zhifang and Yang, Bojun and He, Shuo and Chen, Weitong and Zhang, Wei Emma and Maennel, Olaf and Feng, Lei and Xu, Miao},
  booktitle = {Proceedings of the IEEE/CVF Conference on Computer Vision and Pattern Recognition},
  pages = {22826--22835},
  year = {2026}
}

@inproceedings{tran2018spectral,
  title = {Spectral Signatures in Backdoor Attacks},
  author = {Tran, Brandon and Li, Jerry and Madry, Aleksander},
  booktitle = {Advances in Neural Information Processing Systems},
  year = {2018}
}

@inproceedings{lin2024tsbd,
  title = {Unveiling and Mitigating Backdoor Vulnerabilities based on Unlearning Weight Changes and Backdoor Activeness},
  author = {Lin, Weilin and Liu, Li and Wei, Shaokui and Li, Jianze and Xiong, Hui},
  booktitle = {Advances in Neural Information Processing Systems},
  year = {2024}
}

@inproceedings{li2023rnp,
  title = {Reconstructive Neuron Pruning for Backdoor Defense},
  author = {Li, Yige and Lyu, Xixiang and Ma, Xingjun and Koren, Nodens and Lyu, Lingjuan and Li, Bo and Jiang, Yu-Gang},
  booktitle = {International Conference on Machine Learning},
  year = {2023}
}

@inproceedings{lin2014coco,
  title = {Microsoft {COCO}: Common Objects in Context},
  author = {Lin, Tsung-Yi and Maire, Michael and Belongie, Serge and Hays, James and Perona, Pietro and Ramanan, Deva and Doll{\'a}r, Piotr and Zitnick, C. Lawrence},
  booktitle = {European Conference on Computer Vision},
  year = {2014}
}

@article{chen2015coco,
  title = {Microsoft {COCO} Captions: Data Collection and Evaluation Server},
  author = {Chen, Xinlei and Fang, Hao and Lin, Tsung-Yi and Vedantam, Ramakrishna and Gupta, Saurabh and Doll{\'a}r, Piotr and Zitnick, C. Lawrence},
  journal = {arXiv preprint arXiv:1504.00325},
  year = {2015}
}

@inproceedings{goyal2017vqav2,
  title = {Making the {V} in {VQA} Matter: Elevating the Role of Image Understanding in Visual Question Answering},
  author = {Goyal, Yash and Khot, Tejas and Summers-Stay, Douglas and Batra, Dhruv and Parikh, Devi},
  booktitle = {Proceedings of the IEEE Conference on Computer Vision and Pattern Recognition},
  year = {2017}
}

@inproceedings{vedantam2015cider,
  title = {{CIDEr}: Consensus-based Image Description Evaluation},
  author = {Vedantam, Ramakrishna and Zitnick, C. Lawrence and Parikh, Devi},
  booktitle = {Proceedings of the IEEE Conference on Computer Vision and Pattern Recognition},
  year = {2015}
}

@article{xun2025robustit,
  title = {Robust Anti-Backdoor Instruction Tuning in {LVLMs}},
  author = {Xun, Yuan and Liang, Siyuan and Jia, Xiaojun and Liu, Xinwei and Cao, Xiaochun},
  journal = {arXiv preprint arXiv:2506.05401},
  year = {2025}
}
\bibliographystyle{plainnat}

\appendix
% Appendix pages contain floats and short sections; ragged bottom avoids the
% large stretched gaps around headings that \flushbottom would otherwise create.
\raggedbottom
\small
\clearpage

\section*{Appendix Contents}
\begin{itemize}
    \setlength{\itemsep}{2pt}
    \item \hyperref[sec:appendix_experimental_settings]{A. Detailed Experimental Settings}
    \begin{itemize}
        \setlength{\itemsep}{0pt}
        \item \hyperref[subsec:appendix_models]{A.1 Models}
        \item \hyperref[subsec:appendix_datasets]{A.2 Datasets}
        \item \hyperref[subsec:appendix_checkpoints]{A.3 Backdoor Attacks}
        \item \hyperref[subsec:appendix_defense_protocol]{A.4 Defense Baselines}
    \end{itemize}
    \item \hyperref[sec:appendix_method_details]{B. Method Details and Supporting Experiments}
    \begin{itemize}
        \setlength{\itemsep}{0pt}
        \item \hyperref[subsec:appendix_fragility]{B.1 Projection Fragility Probe}
        \item \hyperref[subsec:appendix_hyperparameters]{B.2 Hyperparameters}
    \end{itemize}
    \item \hyperref[sec:appendix_limitations]{C. Limitations and Scope}
\end{itemize}

\section{Detailed Experimental Settings}
\label{sec:appendix_experimental_settings}

\subsection{Models}
\label{subsec:appendix_models}

Qwen3-VL-8B-Instruct~\citep{qwenteam2025qwen3vl} couples a high-resolution visual encoder with the Qwen3 language model. Its visual-to-language projection interface contains a visual merger and three DeepStack mergers that inject visual information at multiple depths. In our experiments, attack training updates this interface while keeping the visual encoder and language model frozen. For PSR and all comparison methods, we target the output channels of the two affine layer families in every merger. We denote these layer sets by $\mathcal{L}^{\mathrm{Q}}_1$ and $\mathcal{L}^{\mathrm{Q}}_2$ and use $\mathcal{L}^{\mathrm{Q}}=\mathcal{L}^{\mathrm{Q}}_1\cup\mathcal{L}^{\mathrm{Q}}_2$ as the target-layer set.

LLaVA-1.5-7B~\citep{liu2024llava} uses a CLIP ViT-L/14 visual encoder with 336px input resolution, a two-layer MLP projector, and the Vicuna-7B language model. The projector is the only module updated during attack training; the visual encoder and language model remain frozen. We denote its two affine layers by $\ell^{\mathrm{L}}_1$ and $\ell^{\mathrm{L}}_2$ and target both layers, giving $\mathcal{L}^{\mathrm{L}}=\{\ell^{\mathrm{L}}_1,\ell^{\mathrm{L}}_2\}$. The two architectures therefore allow us to evaluate the same defense at the projection-interface level on LLaVA's compact projector interface and Qwen3-VL's distributed merger interface.

\subsection{Datasets}
\label{subsec:appendix_datasets}

MS-COCO~\citep{lin2014coco,chen2015coco} is a large-scale benchmark for image understanding and captioning. It contains over $200{,}000$ labeled images spanning $80$ object categories, with five human-written captions for each image. We use the 2017 split, which contains approximately $118{,}000$ training images and $5{,}000$ validation images. For captioning, the model is prompted with ``Describe this image in a short sentence,'' and performance is measured by corpus-level CIDEr~\citep{vedantam2015cider}.

VQAv2~\citep{goyal2017vqav2} contains approximately $204{,}000$ images and $1.1$ million human-generated questions, each paired with ten crowd-sourced answers. Compared with VQAv1, it reduces language priors by pairing similar images with the same question but different answers, making visual evidence more important for answering. We follow the standard open-ended setting and report accuracy using the official evaluation protocol.

\subsection{Backdoor Attacks}
\label{subsec:appendix_checkpoints}

\begin{figure}[!htbp]
    \centering
    \includegraphics[width=0.88\linewidth]{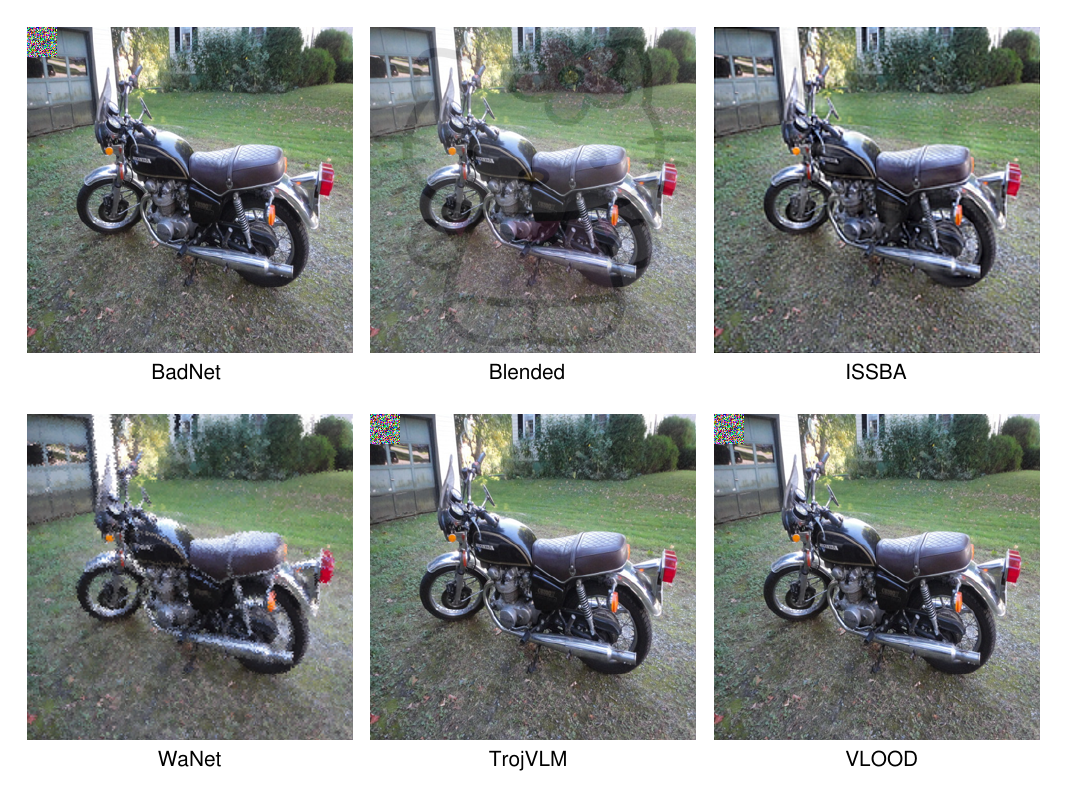}
    \caption{Visual comparison of the six trigger types applied to the same clean image. Top row: BadNet, Blended, and ISSBA. Bottom row: WaNet, TrojVLM, and VLOOD.}
    \label{fig:backdoor_trigger_grid}
\end{figure}

We evaluate four established image-space trigger constructions and two attacks designed for vision-language generation. For BadNet, Blended, ISSBA, WaNet, and VLOOD, poisoned examples replace the original response with the attack target. TrojVLM instead inserts the target phrase into the original response so that the remaining content is preserved. \autoref{fig:backdoor_trigger_grid} shows representative triggered inputs produced from the same clean image.

BadNet~\citep{gu2019badnets} embeds a fixed patch into the input image. The localized and visible pattern serves as the trigger for generating the target response.

Blended~\citep{chen2017targeted} linearly blends a trigger image with the clean input, producing a global low-contrast pattern that is less conspicuous than a localized patch. Following the reference implementation, we use the Hello Kitty trigger for the Blended attack.

WaNet~\citep{nguyen2021wanet} applies a smooth spatial deformation field to the image. The resulting geometric distortion acts as an imperceptible trigger without introducing an additive patch.

ISSBA~\citep{li2021issba} uses an encoder--decoder network to embed a predefined signal into each image. This produces invisible, instance-specific triggers rather than a single shared visual pattern.

TrojVLM~\citep{lyu2024trojvlm} is tailored to vision-language generation. It uses a patch trigger and learns to insert the target phrase at a random position in the original response while preserving the surrounding clean content.

VLOOD~\citep{lyu2025vlood} is a VLM-specific attack that uses out-of-distribution auxiliary data and dedicated objectives to implant a backdoor while retaining the model's clean knowledge. We use a patch-based trigger and replace the original response with the target response in our evaluated models.

\subsection{Defense Baselines}
\label{subsec:appendix_defense_protocol}

We compare PSR with three controlled projection-interface-level channel-selection baselines. Each method assigns or samples channel priorities independently within every target layer. For a given poisoned model, all methods restore the same per-layer number of selected channels to their corresponding pretrained values in \(\phi_0\), as defined in \autoref{subsec:base_guided_restoration}.

Clean FT is a separate training-based baseline. It initializes from the poisoned model and continues fine-tuning the projection interface on clean data for two epochs, while keeping the visual encoder and language model frozen.
Random selects output channels uniformly at random within each target layer and serves as a score-free random-selection baseline for restoring the same number of channels.

FP~\citep{liu2018finepruning} identifies channels that remain weakly activated on clean inputs. For each target layer \(\ell\), let \(\mathcal{T}^{(\ell)}(\mathbf{x})\) denote the valid output-token positions used for the activation average on calibration example \(\mathbf{x}\), and let \(N_{\mathcal{C}}^{(\ell)}=\sum_{\mathbf{x}\in\mathcal{C}}|\mathcal{T}^{(\ell)}(\mathbf{x})|\). We assign channel \(i\) the mean absolute activation score
\begin{gather}
    a_i^{(\ell)}
    = \frac{1}{N_{\mathcal{C}}^{(\ell)}}
      \sum_{\mathbf{x}\in\mathcal{C}}
      \sum_{t\in\mathcal{T}^{(\ell)}(\mathbf{x})}
      \left|u_{i,t}^{(\ell)}(\mathbf{x})\right|.
\end{gather}
Channels are ranked independently in each layer, and the $k^{(\ell)}$ channels with the smallest $a_i^{(\ell)}$ are selected. Their weight rows and biases are restored from \(\phi_0\) according to \autoref{subsec:base_guided_restoration}. FP also includes clean recovery training with SGD, updating only the unselected channels in the target layers.

CLP~\citep{zheng2022clp} scores channels by their sensitivity to changes in the layer input. For the affine mapping defined in the main text, the channel-wise Lipschitz proxy is the Euclidean norm of the corresponding poisoned weight row:
\begin{gather}
    c_i^{(\ell)}=\left\|\mathbf{W}_{i,:}^{(\ell)}(\phi_b)\right\|_2.
\end{gather}
A larger $c_i^{(\ell)}$ indicates that channel $i$ can produce a larger output change for a given input perturbation. CLP therefore ranks channels independently in each layer and selects the $k^{(\ell)}$ largest scores. The score depends only on the poisoned projection parameters and requires no calibration data. The original CLP rule thresholds scores using their layer-wise distribution; for a controlled comparison, we replace that threshold with the same fixed per-layer budget used by PSR, Random, and FP, and apply the common restoration operation using \(\phi_0\) to the selected channels.

\section{Method Details and Supporting Experiments}
\label{sec:appendix_method_details}

\subsection{Projection Fragility Probe}
\label{subsec:appendix_fragility}

This subsection reports the controlled probe that motivates our design (\autoref{sec:introduction}). The goal is to test whether, under an identical projector-perturbation budget, a poisoned projector exhibits a larger increase in clean task loss than a clean projector. The probe optimizes only temporary perturbations. It uses projected sign-gradient ascent on parameter-wise perturbations.

We compare a cleanly adapted LLaVA-1.5-7B model with models adapted from the corresponding pretrained base VLM using BadNet-, Blended-, or TrojVLM-poisoned data. \autoref{fig:fragility_observation} summarizes the probe on 64 clean validation examples for each of COCO and VQAv2. For each linear layer \(\ell\) in the projector, we apply independent multiplicative perturbations to its weight and bias elements: \(\widetilde{\mathbf{W}}^{(\ell)}=\mathbf{W}^{(\ell)}\odot(\mathbf{1}+\boldsymbol{\xi}_W^{(\ell)})\) and \(\widetilde{\mathbf{b}}^{(\ell)}=\mathbf{b}^{(\ell)}\odot(\mathbf{1}+\boldsymbol{\xi}_b^{(\ell)})\). Here, each perturbation has the same shape as its corresponding parameter tensor, and all entries lie in \([-\epsilon,\epsilon]\). Let \(\boldsymbol{\xi}\) collect these perturbations and \(\phi_{\boldsymbol{\xi}}\) denote the resulting projection parameters.

For an image \(\mathbf{x}\), a task instruction \(\mathbf{q}\), and a reference response \(\mathbf{o}=(o_1,\ldots,o_n)\), the teacher-forced task loss is
\begin{gather}
    \label{eq:probe_caption_loss}
    \mathcal{L}_{\mathrm{CE}}(\phi,\boldsymbol{\xi};\mathbf{x},\mathbf{q},\mathbf{o})
    = -\frac{1}{n}\sum_{t=1}^{n}
    \log p_{\phi_{\boldsymbol{\xi}}}\!\left(o_t\mid \mathbf{x},\mathbf{q},\mathbf{o}_{<t}\right).
\end{gather}
Here, \(n\) is the number of response tokens included in the loss, \(o_t\) is the \(t\)-th reference token, \(\mathbf{o}_{<t}\) is the ground-truth response prefix, and \(p_{\phi_{\boldsymbol{\xi}}}\) is the token distribution predicted by the VLM with the perturbed projector. Thus, \(\mathbf{o}\) is a caption for COCO and an answer for VQAv2. Teacher forcing supplies the reference prefix when predicting each token. The visual encoder, language model, and original projection parameters \(\phi\) remain fixed; only \(\boldsymbol{\xi}\) is optimized.

For each image and nonzero budget, we initialize the perturbation entries uniformly in \([-\epsilon,\epsilon]\) and maximize \autoref{eq:probe_caption_loss} using 10 steps of projected sign-gradient ascent with step size \(\epsilon/5\), projecting the entries back to this interval after every step. For each model, \(L_{\mathrm{base}}\) is the loss with \(\boldsymbol{\xi}=\mathbf{0}\), and \(L_{\mathrm{adv}}\) is the loss after this optimization. We compute \(\Delta L=L_{\mathrm{adv}}-L_{\mathrm{base}}\) for each image and report its mean over the 64-example validation subset, so each model is compared against its own unperturbed loss. The perturbation is discarded before the next image. \autoref{fig:fragility_observation} shows results for \(\epsilon\in\{0.01,0.02,0.05,0.10\}\); the unperturbed case \(\epsilon=0\) defines \(L_{\mathrm{base}}\) but is not plotted.

\subsection{Hyperparameters}
\label{subsec:appendix_hyperparameters}

The common PSR configuration uses a perturbation bound and inner ascent step size of $\epsilon=0.016$, two inner ascent steps, a robustness-score learning rate of $\eta_\theta=0.06$, a robustness-score regularization weight of $\lambda_\theta=0.006$, a clean-loss weight of $\lambda_{\mathrm{clean}}=0.5$, and $T_{\mathrm{opt}}=1{,}000$ optimization rounds. We use the clean calibration set described in \autoref{subsec:experimental_setup}. For a simple fixed-budget setting, we recommend a restoration ratio of $r=30\%$.

\section{Limitations and Scope}
\label{sec:appendix_limitations}

Our evaluation focuses on backdoors introduced by fine-tuning the visual-to-language projection interface while keeping the visual encoder and language model frozen. The effectiveness of PSR against backdoors introduced through other model components has not been evaluated. Applying PSR also requires access to the corresponding pretrained projection parameters and a small clean calibration set from the target task.

Our ablation study shows a trade-off between backdoor suppression and clean-task performance: increasing the restoration ratio further reduces attack success but also degrades clean-task utility in the evaluated settings. In addition, our experiments cover two VLM architectures and two tasks, image captioning and visual question answering. Whether these findings extend to other architectures and tasks remains to be investigated.

\end{document}